\documentclass[12pt]{iopart}

\usepackage{graphicx,iopams,bm}

\def\graphicscale{0.32}

\begin{document}

\title[Critical dynamics in trapped particle systems]
{Critical dynamics in trapped particle systems} 

\author{Gianluca Costagliola and Ettore Vicari}

\address{Dipartimento di Fisica dell'Universit\`a di Pisa and 
  INFN, Sezione di Pisa, Largo Bruno Pontecorvo 2, I-56127 Pisa, Italy} 

\date{June 30, 2011}

\begin{abstract}
  We discuss the effects of a trapping space-dependent potential on
  the critical dynamics of lattice gas models.  Scaling arguments
  provide a dynamic trap-size scaling framework to describe how
  critical dynamics develops in the large trap-size limit.  We present
  numerical results for the relaxational dynamics of a two-dimensional
  lattice gas (Ising) model in the presence of a harmonic trap, which
  support the dynamic trap-size scaling scenario.
\end{abstract}




Statistical systems are generally inhomogeneous in nature, while
homogeneous systems are often an ideal limit of experimental
conditions.  Critical phenomena, characterized by the development of
correlations with diverging length and time scales, are usually
considered assuming homogeneous systems.  However, the emergence of
critical correlations is also observed in inhomogeneous systems.
Particularly interesting physical systems are interacting particles
constrained within a limited region of space by an external potential.
This is a common feature of the experimental realizations of
Bose-Einstein condensations in diluted atomic vapors~\cite{CWK-02} and
of optical lattices of cold atoms~\cite{BDZ-08}.  Experimental
evidence~\cite{DRBOKS-07} of a critical behavior has been reported for
a three-dimensional trapped Bose gas, observing an increasing
correlation length compatible with a continuous transition, although
the system is made inhomogeneous by the confining force.  Experimental
evidences of finite-$T$ Kosterlitz-Thouless transitions~\cite{KT-73}
in two-dimensional trapped Bose atomic gases have been provided in
\cite{HSBBD-06,KHD-07,HKCRD-08,CRRHP-09,HZGC-10}.  However, the
inhomogeneity due to the trapping potential is expected to strongly
affects the phenomenology of continuous phase transitions observed in
the absence of a trap.  For example, correlation functions of the
critical modes are not expected to develop a diverging length scale in
a trap.  Therefore, a theoretical description of the critical
correlations in systems subject to confining potentials, and of how
they unfold approaching the transition point, is of great importance
for experimental investigations of the critical behavior of systems of
trapped interacting particles.

We consider a trapping potential
\begin{equation}
V(r) = v^p |\vec{r}|^p \equiv (|\vec{r}|/l)^p,
\label{potential}
\end{equation}
where $v$ and $p$ are positive constants and $l = v^{-1}$ is the {\em
  trap size}, coupled to the particle number.  Harmonic potentials,
  i.e., $p=2$, are usually realized in experiments.  The effect of the
  trapping potential is to effectively vary the local value of the
  chemical potential, so that the particles cannot run away.  Let us
  consider the case in which the system parameters, such as
  temperature, pressure and chemical potential, are tuned to values
  corresponding to the critical regime of the unconfined system.  The
  critical behavior gets distorted by the trap, although it gives rise
  to universal effects in the large trap-size limit, controlled by the
  universality class of the phase transition of the unconfined system.
  The corresponding scaling regime can be cast in the form of a
  trap-size scaling (TSS)~\cite{CV-09,CV-10}, resembling the
  finite-size scaling theory for homogeneous systems, but
  characterized by a further nontrivial {\em trap} critical exponent
  $\theta$, which describes how the length scale $\xi$ of the critical
  modes depends on the trap size at criticality, i.e., $\xi\sim
  l^\theta$, and which can be estimated using renormalization-group
  (RG) arguments.

In this paper we investigate how critical dynamics develop in trapped
systems at thermal continuous transitions, i.e., {\em classical}
transition driven by thermal fluctuations. We extend the TSS Ansatz
of \cite{CV-09}, to allow for the time dependence of the critical
modes, which depends on the static universality class, but also on the
general features of the dynamics, analogously to homogeneous
systems~\cite{HH-77}.

In a standard scenario for a continuous transition, see, e.g.,
~\cite{PV-02}, the
critical behavior of a $d$-dimensional system is characterized by two
relevant parameters $u_t$ and $u_h$, which may be associated with the
temperature $T$, i.e., $u_t\sim T/T_c-1$, and the external field $h$
coupled to the order parameter, whose RG
dimensions are $y_t=1/\nu$ and $y_h= (d+2-\eta)/2$.  The presence of a
trap of size $l$ generally induces a further length scale $\xi_{\rm
trap}\sim l^\theta$, which must be taken into account to describe the
critical correlations.  Within the TSS
framework~\cite{CV-09,CV-10}, see also
\cite{Burkhardt-82,PKT-07,ZD-08,EIP-09,CKT-09,QSS-10,PPS-10,CV-10-e,CV-11}, 
the scaling law of the
singular part of the free energy density around the center of the trap
can be written as
\begin{equation}
F_{\rm sing} = l^{-\theta d} {\cal F}(u_t l^{\theta y_t},
u_h l^{\theta y_h},xl^{-\theta})
\label{freee}
\end{equation}
where $\theta$ is the {\em trap exponent}.  At the critical point
$u_t=0$, the TSS of the correlation function of the order parameter
$\phi$ is
\begin{equation}
G(x,y) \equiv \langle \phi(x) \phi(y) \rangle_c 
= l^{-\theta(d-2+\eta)} f(xl^{-\theta},yl^{-\theta}),
\label{twopf}
\end{equation}
thus implying $\xi\sim l^\theta$ for its length scale.  Finite size
effects, due to a finite volume $L^d$, can be taken into account by
adding a further dependence on $L l^{-\theta}$ in the above scaling
Ansatz~\cite{QSS-10}.

The universal scaling behavior of the critical dynamics is essentially
determined by the dynamic universality class, which depends on a few
general properties of the dynamics, for example, whether there are
conserved quantities during the time evolution (see \cite{HH-77}
for a list of dynamic universality classes).  The scaling behavior 
of the correlation functions requires a further
dynamic critical exponent $z$, which provides the power-law relation
between the critical time scale $\tau$ and the diverging correlation
length $\tau \sim \xi^z \sim u_t^{-z\nu}$, thus implying a
diverging time scale at the critical point.  For example, the
equilibrium time dependence of the correlation function of the order
parameter $\phi(x,t)$ is expected to scale as
\begin{equation}
G(x_1,x_2;t_1,t_2) \equiv \langle\phi(x_1,t_1)\phi(x_2,t_2)\rangle_c
= \xi^{-(d-2+\eta)} f[(x_2-x_1)/\xi,(t_2-t_1)/\xi^z]
\label{gxt}
\end{equation}
Again, the presence of the trap drastically affects the critical
dynamics; in particular, the time correlations are not expected to
develop a diverging time scale.  Within the TSS framework, the time
dependence of the dynamic scaling behavior is expected to
enter through a further
dependence on the scaling variable $tl^{-\theta z}$.  Thus, at the
critical point $u_t=0$, the scaling behavior at equilibrium of the
two-point correlation function (\ref{gxt}) is expected to be
\begin{equation}
G(x_1,x_2;t_1,t_2) = l^{-\theta(d-2+\eta)}
f_g[x_2l^{-\theta},x_1l^{-\theta}, (t_2-t_1)l^{-\theta z}]
\label{gxtt}
\end{equation}
which implies that the corresponding time scale $\tau$ behaves as
\begin{equation}
\tau \sim l^{\theta z}
\label{taul}
\end{equation}
at the critical point.  Analogous scaling behaviors have been put
forward to describe the temperature and time dependence at $T=0$
quantum transitions~\cite{CV-10,CK-10,CV-10-off}.  The dynamic TSS
Ansatz can also be extended to take into account off-equilibrium
dynamics.

In order to check the dynamic TSS scenario, we consider the lattice
gas model defined by the Hamiltonian
\begin{equation}
{\cal H}_{\rm Lgas} = 
- 4 J \sum_{\langle ij\rangle}\rho_i \rho_j - \mu \sum_i \rho_i 
+ \sum_i 2 V(r_i) \rho_i
\label{latticegas}
\end{equation}
where the first sum runs over the nearest-neighbor sites of a square
lattice, $\rho_i=0,1$ depending if the site is empty or occupied by
the particles, $\mu$ is the chemical potential, and $V(r)$ is the
potential (\ref{potential}) which vanishes at the center of the trap.
Far from the origin the potential diverges, thus the expectation value
of the particle number tend to vanish, indeed $\langle \rho_x \rangle
\sim \exp[-2V(x)]$ at large distance from the center, and therefore
the particles are trapped.  The lattice gas model (\ref{latticegas})
can be exactly mapped to a standard Ising model, by replacing
$\rho_i=(1-s_i)/2$, obtaining
\begin{equation}
{\cal H} = - J \sum_{\langle ij\rangle}s_i s_j - h \sum_i s_i 
- \sum_i V(r_i) s_i
\label{ising}
\end{equation}
where $s_i=\pm 1$ and
$h=8J+\mu/2$. In this picture the external potential plays the role
of a space-dependent magnetic field.  In the absence of the trap, the
square-lattice Ising model (\ref{ising}) shows a critical behavior
characterized by the critical exponents $\nu=1$ and $\eta=1/4$, at the
critical point $T=T_c=2/\ln(\sqrt{2}+1)$ (we set $J=1$) and $h=h_c=0$.
Critical correlations do not develop a diverging length scale in the
presence of the external space-dependent potential, i.e., at fixed
$v>0$. But a critical behavior develops in the limit $v\to 0$,
described by TSS as in Eqs.~(\ref{freee}) and (\ref{twopf}). The
trap exponent is given by
\begin{equation}
\theta=16/31, 
\label{thetaval}
\end{equation}
which can be derived
by RG scaling arguments~\cite{CV-09} taking
into account the fact that the external potential couples to the order
parameter.  The static TSS has been numerically investigated
in \cite{CV-09}, here we focus on the dynamic TSS scenario of
Eqs.~(\ref{gxtt}) and (\ref{taul}).

We consider a purely relaxational dynamics (also known as model A of
critical dynamics~\cite{HH-77}), which can be realized by Metropolis
updatings in Monte Carlo simulations.  The corresponding dynamic
exponent $z$ has been accurately determined in homogeneous systems by
numerical equilibrium and off-equilibrium methods, obtaining $z\approx
2.17$.~\footnote{Some of the most recent
equilibrium estimates are $z=2.1667(5)$ from \cite{NB-00},
$z=2.168(5)$ from \cite{WH-97}, $z=2.1665(12)$ from
\cite{NB-96}, $z=2.172(6)$ from \cite{G-95}.
Off-equilibrium results are $z=2.156(2)$ from \cite{SAF-02},
$z=2.153(2)$ from \cite{ZWGY-99}, $z=2.16(3)$ from
\cite{SSS-96}, $z=2.1337(41)$ from \cite{LSZ-95},
$z=2.160(5)$ from \cite{LHAS-95}, $z=2.165(10)$ from
\cite{Ito-93}. The apparent small discrepancy between the two
sets of results should not be significant.}  According to the dynamic
TSS scenario, in the presence of the trap we should not observe any
diverging time scale $\tau$, but the critical dynamics is only
recovered in the limit $l\to \infty$, with a power-law diverging time
scale (\ref{taul}).

In our simulations we consider square lattices with $-L \le x,y \le
L$, and trap potential $V(x) = (x^2+y^2)/l^2$ (thus the trap is
centered at the origin), while at the boundaries $|x|,|y|=L+1$ the
spin variable is kept fixed: $s_i=1$ (corresponding to $\rho_i=0$).
In the presence of the trap of size $l$, the lattice size $L$ is
chosen to have the spin variables effectively frozen at the boundary,
making unnecessary the use of larger lattices to effectively obtain
infinite volume results.  See also below.  The dynamics is provided by
Metropolis updatings of the lattice variables using a checkerboard
scheme (a time unit corresponds to a sweep of the lattice).  We
present results of MC simulations for several values of the trap size,
up to $l=320$. The typical statistics of our simulations
is $10^9$ sweeps.

We want to determine the power-law behavior of the time scale at
$T=T_c$ as a function of the trap size. For this purpose we estimate
the autocorrelation times of the magnetization along the lattice axes,
which can be easily computed. More precisely, we define
\begin{equation}
M_r \equiv \sum_{i=-\lfloor rl^\theta\rfloor}^{\lfloor rl^\theta\rfloor} 
s_{i0}+s_{0i}
\label{mrdef}
\end{equation}
where $\lfloor x \rfloor\equiv{\mathop{\rm floor}}(x)$ is the largest
integer not greater than $x$, the distance $r$ is measured in unit of
the length scale $l^\theta$ (to construct nonlocal observables
compatible with the expected TSS behavior), and the subscripts of
$s_{ij}$ indicate the coordinate of the lattice sites.  In the
following analysis we consider the values $r=1/2$ and $r=1$.  We
estimate the autocorrelation times of the correlators
\begin{equation}
C_r(t_2-t_1) = \langle M_r(t_1)M_r(t_2) \rangle_c 
\label{crdef}
\end{equation}
For this purpose we exploit the method of \cite{HPV-07} which
provides a substantial numerical advantage because it does not require
extrapolated large-time estimates of the autocorrelation functions.
We define
\begin{equation}
\hat\tau_r(t+n/2) \equiv {n\over \ln [C_r(t)/C_r(t+n)]},
\label{defBL}
\end{equation}
where $n$ is a fixed integer number. A linear interpolation can be
used to extend $\hat\tau_r(t)$ to all real numbers.  Then, for any
positive $x$, we define an autocorrelation time $\tau_{r,x}$ as the
solution of the equation~\footnote{ This definition is based on the
idea that, if the autocorrelation function $C(t)$ were a pure
exponential, i.e., $C(t) = C_0 \exp(-t/\tau)$, then $\hat\tau(t) =
\tau$ for all $t$ and thus $\tau_x = \tau$ for any $x$.  It
provides a good autocorrelation time for any $x$, which
converges to the integrated autocorrelation time $\tau_{{\rm int}} =
{1\over2} \sum_{t=-\infty}^\infty \, C(t)/C(0)$ for $x\to
\infty$.  See \cite{HPV-07} for a through discussion.}
\begin{equation} 
\tau_{r,x} = \hat\tau_r(x\tau_{r,x}).
\label{tauxdef}
\end{equation}
$\tau_{r,x}$ are expected to show the same power-law behavior with
increasing $l$, i.e., $\tau_{r,x}\sim l^\kappa$.  After some checks,
we chose $x=2$ as optimal value for our calculations, so in the
following we present only results for $x=2$ and skip the subscript
indicating the value of $x$.

\begin{figure}[tbp]
\begin{center}
\includegraphics*[scale=\graphicscale]{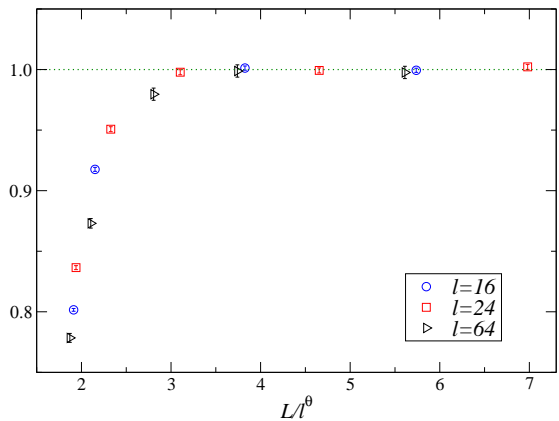}
\end{center}
\vskip-5mm
\caption{ The ratio $\tau_{1/2}(L)/\tau_{1/2}(\infty)$ 
vs $L l^{-\theta}$.  }
\label{figfss}
\end{figure}

Let us first discuss the finite size effects due to a finite volume
$L^2$, which are expected to enter through the FSS variable $L
l^{-\theta}$~\cite{QSS-10}.  This is confirmed by Fig.~\ref{figfss},
which shows the data of the ratio $\tau_{1/2}(L)/\tau_{1/2}(\infty)$
between the autocorrelation times of $M_{1/2}$ at a given $L$ and in
the $L\to\infty$ limit, as estimated by the result for the largest
available lattice. The analysis of the FSS effects show that, within
an accuracy of a few per mille, the data for $Ll^{-\theta}\gtrsim 5$
can be effectively considered as $L\to\infty$ data.

\begin{figure}[tbp]
\begin{center}
\includegraphics*[scale=\graphicscale]{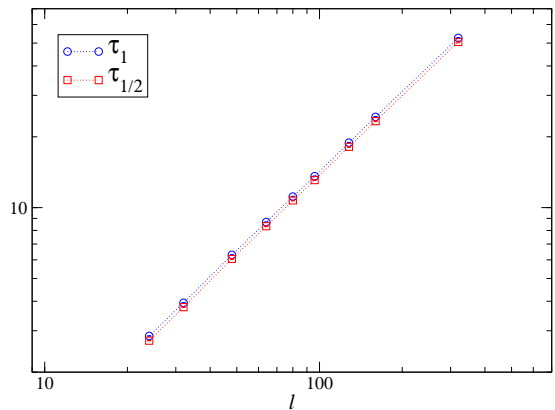}
\end{center}
\vskip-5mm
\caption{ Power-law behavior of the autocorrelation times $\tau_{1/2}$
and $\tau_{1}$.  The lines are drawn to guide the eyes.  }
\label{figtau}
\end{figure}

In Fig.~\ref{figtau} we show the resulting infinite-volume estimates
of the autocorrelation times of $M_{1/2}$ and $M_{1}$, up to a trap
size $l=320$.  They clearly show an asymptotic power-law behavior.
The data for $l\ge 48$ turn out to fit the simple power law
$\tau_r=al^\kappa$ with $\chi^2/{\rm d.o.f}\lesssim 1$, providing the
estimate $\kappa=1.120(3)$ (the results of the fits appear reasonably
stable with increasing the minimum value of $l$ allowed in the fit,
while the data for $l<48$ show small deviations from the simple power
law).  Analogous results are obtained by varying the value of $x$ in
the definition (\ref{tauxdef}).  Dividing the exponent $\kappa$ by
$\theta=16/31$, we obtain
\begin{equation}
\kappa/\theta= z =2.170(6)
\label{finalest}
\end{equation}
which is in good agreement with the best available estimates of $z$,
for example~\cite{NB-00} $z=2.1667(5)$.  We have also performed fits to
all data including scaling corrections, which are expected to
be~\cite{CV-10} $O(l^{-\theta})$, obtaining consistent results.

In conclusion, our numerical results for the relaxational dynamics of
the square-lattice gas model (\ref{latticegas}) show that the dynamic
TSS scenario correctly describes the development of a critical
dynamics in the presence of an inhomogeneous external potential which
acts as a trap for particle systems. We expect that the validity of
the dynamic TSS extends to other dynamic universality classes.

\end{document}